\documentclass[letter]{aa}
\usepackage{txfonts}
\usepackage[labelsep=period]{caption}

\usepackage{longtable}

\usepackage{makecell}
\usepackage{graphicx}
\usepackage{natbib}
\usepackage[utf8]{inputenc}
\usepackage{natbib}
\usepackage[mathscr]{euscript}
\usepackage{lscape}

\usepackage{booktabs,caption}
\usepackage[flushleft]{threeparttable}

\begin{document}

  \title{Evidence for $\gtrsim{4}$~Gyr timescales of neutron star mergers\\ from Galactic archaeology}
   
   \author{\'{A}.~Sk\'{u}lad\'{o}ttir
   		\inst{1,2}
        \and
       	S.~Salvadori \inst{1,2}
       	}
          
   \institute{	 
   			 Dipartimento di Fisica e Astronomia, Universit\'{a} degli Studi di Firenze, Via G. Sansone 1, I-50019 Sesto Fiorentino, Italy.\\
   			 \email{asa.skuladottir@unifi.it}
   			   \and
   			   INAF/Osservatorio Astrofisico di Arcetri, Largo E. Fermi 5, I-50125 Firenze, Italy.
                }

\abstract{The nucleosynthetic site of the rapid ($r$) neutron-capture process is currently being debated. The direct detection of the neutron star merger GW170817, through gravitational waves and electromagnetic radiation, has confirmed such events as important sources of the $r$-process elements. However, chemical evolution models are not able to reproduce the observed chemical abundances in the Milky Way when neutron star mergers are assumed to be the only $r$-process site and realistic time distributions of such events are taken into account. Now for the first time, we combine all the available observational evidence of the Milky Way and its dwarf galaxy satellites to show that the data can only be explained if there are (at least) two distinct $r$-process sites: a \textit{quick} source with timescales comparable to core-collapse supernovae, $t_\textsl{quick}\lesssim10^8$~yr, and a \textit{delayed} source with characteristic timescales $t_\textsl{delayed}\gtrsim4$~Gyr. The delayed $r$-process source most probably originates in neutron star mergers, as the timescale fits well with that estimated for GW170817. Given the short timescales of the quick source, it is likely associated with massive stars, though a specific fast-track channel for compact object mergers cannot be excluded at this point. Our approach demonstrates that only by looking at all the available data will we be able to solve the puzzle that is the $r$-process.
}


   \maketitle

%
\section{Introduction}

              \begin{figure*} 
   \centering
   \includegraphics[width=\hsize-3cm]{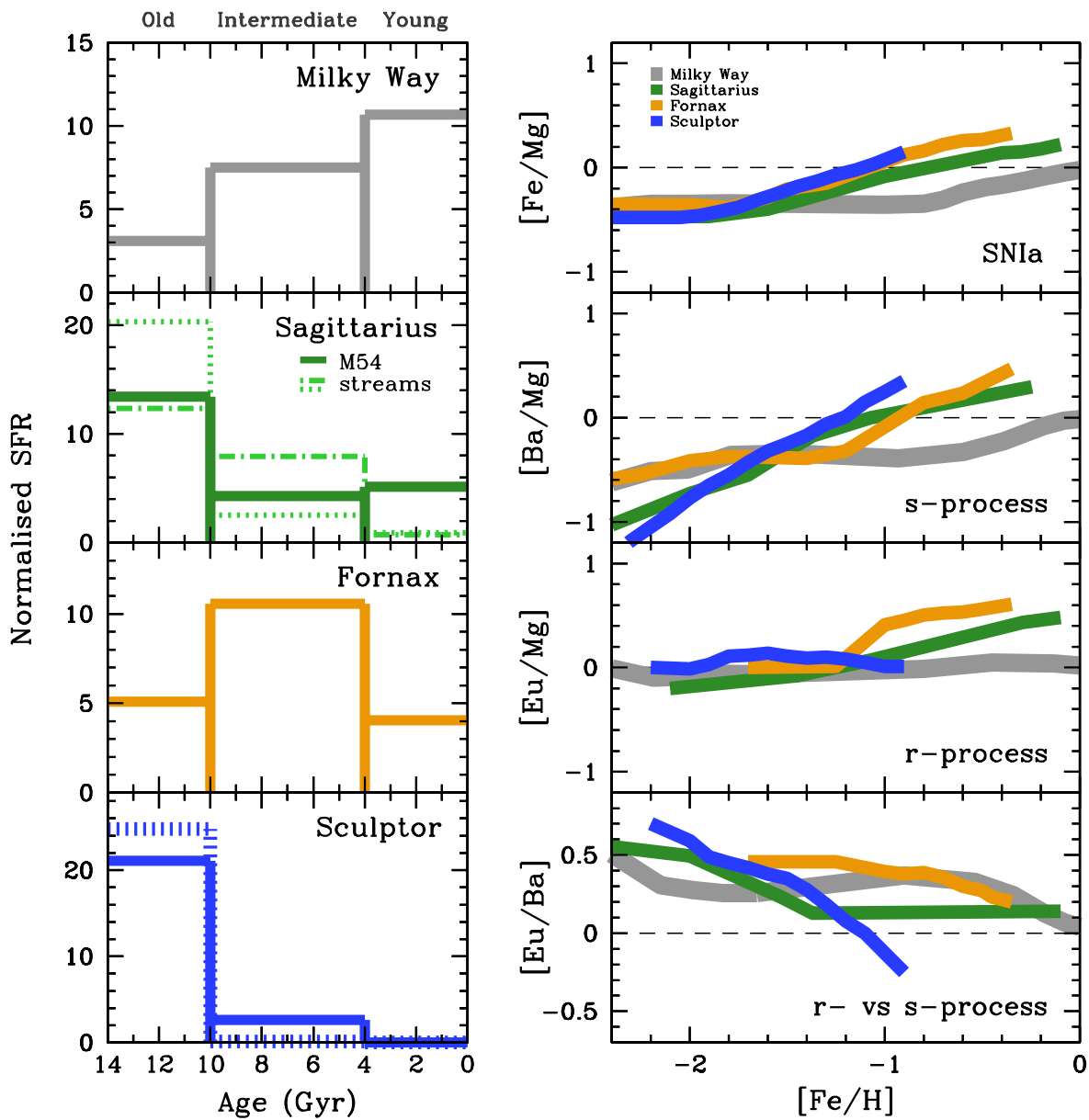}
      \caption{Schematic figure of the SFHs (left) and abundance ratios (right) of the Milky Way solar neighborhood (gray), and the dSph galaxies: Sagittarius (green), Fornax (yellow), and Sculptor (blue). In the case of Sagittarius, SFHs of both the nucleus M54 and two of its streams are shown. Two independent studies of the SFH of Sculptor are included. For references and more details see Sect.~\ref{obsdata}.
      } \label{fig:sfhabu}
   \end{figure*}   

The Laser Interferometer Gravitational-Wave Observatory (LIGO) detection of the event GW170817 is one of the major milestones of modern astrophysics \citep{Abbott17a}. Through observations of gravitational waves and electromagnetic radiation, GW170817 was shown to be the first unambiguous detection of a neutron star merger (NSM), and also proved them to be a source of short gamma-ray bursts (GRBs; \citealt{Abbott17b,Abbott17c}). The electromagnetic radiation following GW170817 has confirmed that NSMs are an important source of the rapid ($r$) neutron ($n$)-capture process \citep{Chornock17,Cowperthwaite17,Drout17,Pian17,Tanaka17,Villar17,Watson19}. Furthermore, GW170817 is currently the only NSM to date with a confirmed detection of its host galaxy, which is an early-type galaxy with a predominantly old stellar population \citep{Abbott17a, Coulter17}. Different analyses have thus concluded long time delay of this event: $>3$~Gyr by \citet{Pan17}; and  6.8-13.6~Gyr by \citet{Blanchard17}.

The long timescales which have been estimated for GW170817 and the fact that a large fraction of short GRBs are associated with early-type galaxies \citep{Berger14,Fong17} pose a problem for Galactic archaeology. The observed stellar abundances of $r$-process elements in the Milky Way have proven difficult to reproduce when NSMs are assumed as the dominant (or only) source, both at the high- and the low-metallicity end. To date, no star has been found without any traces of an $r$-process enrichment, even at $\text{[Fe/H]}<-3$ (e.g.~\citealt{Francois07,Roederer14b}). Thus it is required that NSMs occur very early after the onset of star formation. At higher metallicity, $\text{[Fe/H]}>-1$, the abundances of the $r$-process element\footnote{$\sim$94\% of the Eu solar abundance comes from the $r$-process, according to \citet{Bisterzo14}.} Eu have been proven impossible to reproduce with NSMs as the only $r$-process source when physically motivated time delay distributions are adopted (e.g.~\citealt{Shen15,Voort15,Komiya16,Cote19,Hotokezaka18,Simonetti19}). On the other hand, if short timescales of NSMs are assumed ($\lesssim10^8$~yr), the abundance trend  of the Milky Way disks can be readily explained (e.g.~\citealt{Grisoni19}). However, given what is known of the NSM GW170817, and short GRBs in general \citep{Berger14,Fong17}, such time-delay distributions of NSMs are highly implausible (see also the detailed discussions in \citealt{Skuladottir19} and \citealt{Cote19}).

Attempts have been made to solve this discrepancy between the time-delay distributions of NSMs and Galactic archaeology by invoking specific mechanisms that influence the abundances, such as natal kicks \citep{Tauris17}, metallicity dependence of NSM frequency \citep{Simonetti19}, or details in the physics of the interstellar medium (ISM; \citealt{Schonrich19}). However, all of the proposed mechanisms have to be specifically tailored for the Milky Way abundances, and are thus unable to self-consistently explain the observed Eu abundances in several galaxies at once \citep{Skuladottir19,Bonetti19}. This highlights the importance of including all the available data when proposing scenarios to reconcile studies of NSMs and that of Galactic archaeology.

\section{Observational data}\label{obsdata}

With the goal of better understanding the $r$-process and its nucleosynthetic site(s), we gathered the available chemical abundances of stars in the Milky Way and its three largest satellite dwarf spheroidal (dSph) galaxies: Sagittarius, Fornax, and Sculptor. The aim is to put these chemical abundances in context with the star formation histories (SFHs) of these galaxies.

\subsection{Star formation histories}

A schematic view of the SFHs of different galaxies is shown in the left panel of Fig.~\ref{fig:sfhabu}. To simplify the discussion and to avoid uncertainties in the details of these SFHs, we define three broad age regions for star formation and stellar ages: $\textit{Old}\geq10$~Gyr; $10~\text{Gyr}<\textit{Intermediate}<4$~Gyr; and $\textit{Young}\leq4$~Gyr. These definitions will be used hereafter.

As a proxy for the SFH of the solar neighborhood, we use the astroseismic age measurements of \citet{Silva18}, which include $\approx2000$ stars out to distances of 2~kpc from the Sun. This work corrects for selection effects, and is consistent with the general consensus on the SFH of the solar neighborhood, i.e. a prominent peak in star formation 2-4~Gyr ago (\citealt{Cignoni06,Bernard18,Mor19,Isern19}).

A full SFH of the Sagittarius dSph galaxy is currently unavailable. A recent study by \citet{Alfaro-Cuello19} provides ages of $\approx6600$ stars in the Sagittarius nuclear star cluster, M54, which we use as a proxy of its SFH in Fig.~\ref{fig:sfhabu} (dark green solid line). As in the original work, we only include stars whose ages have less than 40\% uncertainty. 
Sagittarius is currently being disrupted, and multiple stellar streams in the Milky Way halo have been identified as originating from this dSph. The SFH of the Sagittarius streams within the Stripe 82 region were studied by \citet{deBoer15}, and are also shown in Fig.~\ref{fig:sfhabu} for comparison, both for the bright stream (dotted-dashed green line), and the faint stream (dotted green line). For details see \citet{deBoer15}.

The SFH of Fornax is adopted from \citet{deBoer12f}. Two independent studies of the Sculptor SFH are shown in Fig.~\ref{fig:sfhabu}: the solid line shows that of \citet{deBoer12}, and the dashed line is from \citet{Bettinelli19}.

\begin{table}
\caption{Slopes of [Eu/Mg] with [Fe/H] in the Milky Way and its largest dSph satellite galaxies.}
\label{tab:slopes} 
\centering
\begin{tabular}{c c c c}
\hline\hline
Galaxy &	$\Delta\text{[Eu/Mg]}/\Delta\text{[Fe/H]}$	&	[Fe/H] range \rule[-1.2ex]{0pt}{0pt} \rule{0pt}{2.ex}\\  

\hline 
Milky Way & $+0.06\pm0.01$ & $-3\leq\text{[Fe/H]}\leq0$\\
Milky Way & $+0.00\pm0.04$ & $-1\leq\text{[Fe/H]}\leq0$\\
Sagittarius & $+0.43\pm0.05$ &$-2.6\leq\text{[Fe/H]}\leq-0.9$\\
Fornax & $+0.31\pm0.32$ & $-1.7\leq\text{[Fe/H]}\leq-0.3$\\\
Sculptor & $+0.03\pm0.10$ & $-2.2\leq\text{[Fe/H]}\leq-0.9$ \\
\hline
\end{tabular}
\end{table}

\subsection{Chemical abundances}

Schematic chemical abundance trends in the Milky Way, Sagittarius, Fornax and Sculptor are shown in Fig.~\ref{fig:sfhabu} (right panels). To avoid dealing with systematic errors between surveys as much as possible, we only included two surveys of Milky Way chemical abundances: \citet{Roederer14} at $\text{[Fe/H]}\lesssim-1$, and \citet{Mishenina13} at $\text{[Fe/H]}\gtrsim-1$.

The abundances of the Sagittarius dSph shown in Fig.~\ref{fig:sfhabu} are only based on high-resolution (HR) UVES VLT or the Magellan MIKE spectra presented in \citet{Bonifacio00}, \citet{Sbordone07}, \citet{McWilliam13}, and \citet{Hansen18}. The Mg abundances for the \citet{Hansen18} sample come from Sk\'{u}lad\'{o}ttir, Hansen \& Salvadori (in prep.). \citet{Sbordone07} and \citet{McWilliam13} identified their sample as M54 member stars, while the metal-poor stars ($\text{[Fe/H]}<-1$) of \citet{Hansen18} lie outside the virial radius of the nuclear cluster, but are still centrally located in Sagittarius.

Chemical abundances in Fornax come from Reichert et al. (submitted to A\&A). The abundance measurements for Mg, Ba and Eu in Sculptor are derived in \citet{Skuladottir19}, who adopted the stellar parameters and [Fe/H] from \citet{Hill19}.

For all galaxies, outliers such as stars showing signatures of the $s$-process (i.e. high Ba abundances accompanied by high C abundances) are not included when deriving the main trends shown in Fig.~\ref{fig:sfhabu} or the slopes shown in Table~\ref{tab:slopes}. These stars show the chemical enrichment of binary transfer \citep{Lucatello05,Starkenburg14,HansenT16s} and are thus not indicative of the chemical evolution of the ISM. Apart from these outliers, the scatter in different abundance ratios is in most cases compatible with measurement errors. An exception is Fornax whose abundance ratios show a significant scatter, which is revealed in the large uncertainties on the slope in Table~\ref{tab:slopes}. In the case of [Ba/Mg] there is also an increase in scatter at $\text{[Fe/H]}\lesssim-2$, and for [Eu/Mg] at  $\text{[Fe/H]}\lesssim-2.5$ (see also \citealt{Skuladottir19}). This does not affect our conclusions with regard to the $r$-process timescales.

              \begin{figure} 
   \centering
   \includegraphics[width=\hsize-0.5cm]{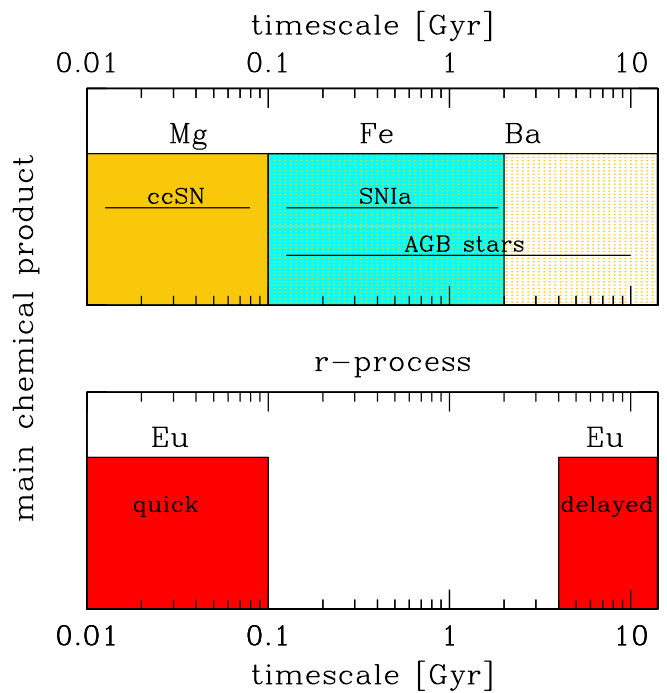}
      \caption{\textit{Top panel:} Schematic timescales of the main nucleosynthetic sites: ccSN, SNIa and AGB stars, each labeled with a reference element. The timescales of ccSN and AGB stars are estimated from stellar lifetimes \citep{Raiteri96}, while that of SNIa refers to the peak of the delay-time distribution function based on data and models \citep{Maoz14}. \textit{Bottom panel:} Our inferred timescales of the quick and delayed $r$-process sources. 
      } \label{fig:nucl}
   \end{figure}   

\section{Constraining the $r$-process timescales}

To investigate the timescales of the $r$-process it is useful to compare its main chemical products (e.g. Eu) to those of other known processes (see Fig.~\ref{fig:nucl}). The chemical abundance ratios of [Fe/Mg], [Ba/Mg], [Eu/Mg], and [Eu/Ba] are therefore shown in the right panels of Fig.~\ref{fig:sfhabu} for the Milky Way and its three largest dSph galaxies, Sagittarius, Fornax, and Sculptor. 

The element Mg is almost exclusively made in core-collapse supernovae (ccSN; e.g. \citealt{Nomoto13}) which have short timescales (see Fig.~\ref{fig:nucl}) and create $\sim$99\% of the solar abundance \citep{Tsujimoto95}. On the other hand, the majority of Fe and Ba in the Sun come from nucleosynthetic channels with significant time delay relative to ccSN. Around $\sim$57\% of Fe in the solar neighborhood is estimated to come from supernovae type~Ia (SNIa; e.g. \citealt{Tsujimoto95}), which have characteristic timescales on the order of $\sim1$~Gyr (e.g. \citealt{Maoz14}; see Fig.~\ref{fig:nucl}). Similarly, $\sim$85\% of the Ba in the Sun comes from the slow ($s$) $n$-capture process \citep{Bisterzo14}, which occurs in asymptotic giant branch (AGB) stars with $m_\star\lesssim8$~M$_\sun$ (e.g.~\citealt{Karakas14}). The delay times therefore depend on the lifetimes of the stars, which range from $\sim10^8$~yr for the most massive stars to the age of the Universe (and longer) for the lowest mass stars (see Fig.~\ref{fig:nucl}).

\subsection{Delayed processes}

Fig.~\ref{fig:sfhabu} shows how both [Fe/Mg] and [Ba/Mg] increase with metallicity in \textit{all} depicted galaxies. This is a clear signature of processes that are delayed relative to ccSN, in this case SNIa and AGB stars, respectively. At the earliest stages in the evolution of any galaxy, its chemical enrichment is dominated by fast sources such as ccSN (see Fig.~\ref{fig:nucl}), which create large amounts of Mg and other $\alpha$-elements, but also some Fe and other iron-peak elements (e.g.~\citealt{Nomoto13}). As time passes, the delayed processes (SNIa and AGB stars) start to contribute to the chemical enrichment, and thus the abundance ratios [Fe/Mg] and [Ba/Mg] start to increase. The increase in [Ba/Mg] in Fig.~\ref{fig:sfhabu} starts at lower metallicity than that of [Fe/Mg]. This is in agreement with the findings of \citet{Amarsi19}, who concluded that the influences of AGB stars in the Milky Way disks become apparent earlier than than those of SNIa. 

The increase in both [Fe/Mg] and [Ba/Mg] occurs at lower metallicity in the dwarf galaxies compared to the Milky Way (see Fig.~\ref{fig:sfhabu}). This is caused by less efficient chemical enrichment of the smaller galaxies, which are only able to enrich their ISM up to low metallicity (and thus low [Fe/H]) before the onset of the delayed processes, compared to the larger, more efficiently star forming Milky Way (e.g. \citealt{Tolstoy09,Hill19}). 

Furthermore, higher values of [Fe/Mg] and [Ba/Mg] are reached in the dwarf galaxies than in the Milky Way (see Fig.~\ref{fig:sfhabu}). Contrary to the Milky Way, none of the dSph galaxies are currently forming stars. The largest dSphs, Sagittarius and Fornax, have not been forming stars for the last $\sim$1-2~Gyr \citep{deBoer12f,Alfaro-Cuello19}, while Sculptor has not formed stars for $\sim$6-10~Gyr \citep{deBoer12,Bettinelli19}, see Fig.~\ref{fig:sfhabu}. As these dSph galaxies lost their gas and star formation ceased, the production of  Mg from ccSN consequently decreased, while the amount of elements released by delayed processes (e.g.~Fe and Ba) was set by a higher star formation rate (SFR) earlier on. These delayed products were then released into a small mass of gas, enhancing their influence \citep{Hill19}. 

Finally, we note that the SFHs and metallicity evolution with time are far more complex than implied by our schematic Fig.~\ref{fig:sfhabu}. In  the dSph galaxies, metallicities are shown to have fairly clear correlations with age \citep{deBoer12,deBoer12f,Hill19,Alfaro-Cuello19}. In the Milky Way, however, this correlation is not very strong (e.g.~\citealt{Salvadori10,Silva18}). Different components of the Milky Way have distinct SFHs where major merger events and stellar migration play a significant role (e.g.~\citealt{Helmi18,Feuillet19}), but increasing [X/Mg] with age is also observed in solar twins for products of delayed processes, such as SNIa and AGB stars (e.g.~\citealt{Nissen18}). In general we can thus conclude that increasing [X/Mg] with [Fe/H] is a clear sign of a process that is delayed relative to ccSN. However, the quantitative evolution of [X/Mg] with [Fe/H] is highly dependent on the SFH of the galaxy and galaxy component in question.

\subsection{Timescales of the $r$-process}

From Fig.~\ref{fig:sfhabu} it is clear that [Eu/Mg] does not behave in the same way as the known delayed processes, which are traced by [Fe/Mg] and [Ba/Mg]. The four galaxies do not give a consistent picture, where the Milky Way and Sculptor have a flat (or almost flat) trend of [Eu/Mg] with [Fe/H], while  in the case of Fornax and Sagittarius there is a clear increase in [Eu/Mg] with metallicity (see also Table~\ref{tab:slopes}). As shown in \citet{McWilliam13}, this increase cannot be caused by the $s$-process (which produces a very small amount of Eu) since in both galaxies $\text{[Eu/Ba]}>0$ (Fig.~\ref{fig:sfhabu}). 
Similarly, these high [Eu/Mg] values in Sagittarius and Fornax cannot come from the intermediate ($i$) $n$-capture process since it also predicts $\text{[Eu/Ba]}<0$ \citep{Hampel16}. Therefore, Sagittarius and Fornax are enhanced in the $r$-process, both with respect to ccSN ($\text{[Eu/Mg]}>0$) and to the $s$-process ($\text{[Eu/Ba]}>0$).

The observational data shown in Fig.~\ref{fig:sfhabu} are therefore inconsistent with only one $r$-process source:

\begin{itemize}

\item One source with short timescales, $\lesssim10^8$~yr, cannot explain the increase in [Eu/Mg] with [Fe/H] in Fornax and Sagittarius.
\item A single source with timescales similar to or longer than SNIa and AGB stars would result in much steeper [Eu/Mg] slopes in Sculptor and the Milky Way. Furthermore, such a source does not explain why $r$-process elements are found in stars at the lowest $\text{[Fe/H]}<-3$, which typically have no sign of either SNIa or AGB enrichment.
\end{itemize}

\noindent This discrepancy can only be resolved by adopting (at least) two $r$-process channels, one \textit{quick} and one \textit{delayed} source, see Fig.~\ref{fig:nucl}.

The quick source would have to be on timescales shorter than AGB stars, i.e. $\lesssim{10^8}$~yr. This quick source is consistent with having the same timescale as ccSN since the slopes of [Eu/Mg] in Sculptor and in the Milky Way disks ($\text{[Fe/H]}<-1$) are consistent with zero (see Table~\ref{tab:slopes}). 

Quantifying the characteristic timescales of the delayed source is more challenging. However, it is immediately clear that the delayed $r$-process source has to have timescales that are longer than that of ~SNIa, otherwise a clear increase in [Eu/Mg] with [Fe/H] would appear in Sculptor and the Milky Way, similar to the increase in [Fe/Mg]. To explain the absence of a signature of the delayed source in Sculptor, it needs to have characteristic timescales that are \textit{longer than the Sculptor SFH.} Fig.~\ref{fig:sfhabu} shows that SFH of Sculptor spans $\sim$4-6~Gyr, depending on the study \citep{deBoer12,Bettinelli19}. We can therefore conclude that a delayed source with $t_\textsl{delayed}\gtrsim4$~Gyr can explain both the abundance pattern observed in the two large dSph galaxies with extended SFHs, Sagittarius and Fornax, and in the smaller Sculptor, which has a truncated SFH (see Fig.~\ref{fig:sfhabu}).

When looking closely at the left panels of Fig.~\ref{fig:sfhabu}, it becomes clear why a delayed $r$-process source, with $t_\textsl{delayed}\gtrsim4$~Gyr, would not significantly affect the chemical evolution of the solar neighborhood. The SFH of the solar neighborhood has a prominent peak around 2-3~Gyr ago \citep{Silva18,Mor19}. This would enrich the ISM with large amounts of Mg from ccSN, and would thus similarly provide a large increase in products from the quick $r$-process source. The quick source would then completely dominate the production of Eu relative to the delayed source, whose influence is determined by a lower SFR earlier on. The time since this peak, 2-3~Gyr, is long enough for us to see the increase in Fe from SNIa and in Ba from AGB stars at $\text{[Fe/H]}>-1$ resulting from this peak of star formation. On the other hand, it is too short for the ISM to be enriched from the delayed $r$-process sites formed in this peak of SFH. Contrary to the solar neighborhood, more than $60\%$ of the stars in Sagittarius \citep{Alfaro-Cuello19} and Fornax \citep{deBoer12f} formed more than 6~Gyr ago, and have SFHs extended enough to show the signatures of these old stars. Thus, the delayed source has much greater influence.

Our proposed scenario of (at least) two $r$-process sites is thus able to consistently explain the observational evidence from the Milky Way and its largest dSph galaxies. Fig.~\ref{fig:nucl}, shows our inferred timescales of the two $r$-process sources, with distinct characteristic timescales, a quick source with $t_\textsl{quick}\lesssim10^8$~yr, and a delayed source with $t_\textsl{delayed}\gtrsim 4$~Gyr.

\section{Testing our scenario}

To ensure that we have a self-consistent picture, it is important to see if other available data is in agreement with our proposed scenario of two distinct $r$-process sites with different characteristic timescales, namely a quick and a delayed source, see Fig.~\ref{fig:nucl}.

\subsection{Solar twins}

If there are two important $r$-process sites, we expect to see signatures of this in the high-quality data of solar twins. Stars with atmospheric parameters and [Fe/H] similar to those of the Sun, i.e. solar twins, allow chemical abundance measurements of unparalleled precision \citep{Nissen18}.

              \begin{figure}
   \centering
   \includegraphics[width=\hsize-1.5cm]{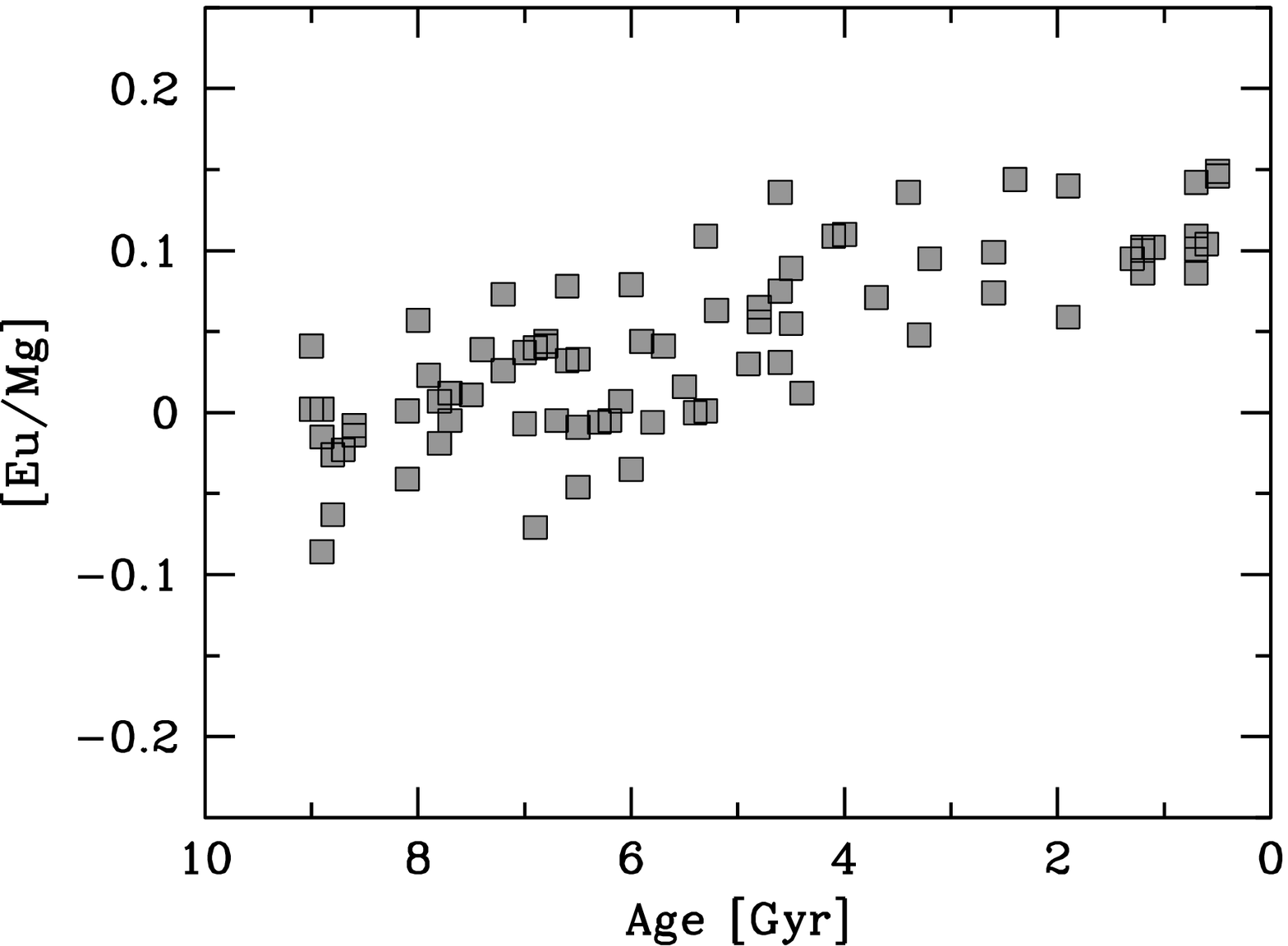}
      \caption{Ratios of [Eu/Mg] in solar twins with different ages (\citealt{Spina18} (ages, Eu); \citealt{Bedell18} (Mg)).
      }
         \label{fig:st}
   \end{figure}

Fig.~\ref{fig:st} shows [Eu/Mg] with age in solar twins with astroseismic age measurements, which are among the most accurate stellar ages achievable to date (e.g.~\citealt{Garcia19}). There is a clear increase in [Eu/Mg] with age, indicative of a delayed component in the production of Eu. However, the slope of [Eu/Mg] with age (see also Table~\ref{tab:st}) flattens out around $\sim4$~Gyr ago, when the solar neighborhood was close to reaching its peak in SFR \citep{Silva18}. This is consistent with our previous discussion, and suggests that the quick source has been the dominant source of the $r$-process over the last $4$~Gyr.

\begin{table}
\caption{Slopes $\beta$ of [Eu/Mg] with age (absolute values) in the sample of solar twins from Fig.~\ref{fig:st}.
}
\label{tab:st} 
\centering
\begin{tabular}{c c}
\hline\hline
Age range &	Slope $\beta$\rule[-1.2ex]{0pt}{0pt} \rule{0pt}{2.ex}\\ 
 &  (dex/Gyr)\\
\hline 
All & 0.0168 $\pm$  0.0015 \\
$\leq$ 4 Gyr & 0.0076 $\pm$ 0.0052 \\
$>4$ Gyr & 0.0175 $\pm$ 0.0034\\
\hline
\end{tabular}
\end{table}

\subsection{Smaller Milky Way dwarf galaxies}

Small dSph galaxies around the Milky Way (i.e. smaller than Sculptor, $M_\star<10^6~$M$_\odot$) have a wide range of [Eu/Mg] distributions: they either increase or decrease with [Fe/H] and typically have a significant scatter (e.g. \citealt{Shetrone01,Shetrone03,Sadakane04,Cohen09,Cohen10,Norris17}). This stochasticity suggests that the rare and prolific events of the $r$-process are not well sampled in these small systems, which likely had a much lower SFR than larger systems (e.g. \citealt{Salvadori15}). If only a handful of events occur in a galaxy with a SFH of several Gyr, chance plays a significant role in the distribution of [Eu/Mg] with [Fe/H] (see further discussion in \citealt{Skuladottir19}).

The ultra-faint dwarf (UFD) galaxies, $M_\star\lesssim10^5$, typically have either very low $\text{[Eu/Mg]}<0$ (e.g. \citealt{Frebel10UFD,Frebel14,RoedererKirby14,Ji19ApJ}) or extremely high $\text{[Eu/Mg]}>+0.5$ (e.g.~\citealt{Ji16Nat,Ji16ApJ}). This is consistent with those small systems having experienced one $r$-process event or none. The SFHs of UFDs typically only span 1-2~Gyr \citep{Salvadori09,Salvadori15,Brown14}. This does not provide any tight constraints on the timescales of the $r$-process. If we assume two $r$-process sites, one quick and one delayed, a single $r$-process event happening in an UFD can either belong to the quick source, or, alternatively to a rare early occurrence of the delayed source.

These small individual systems, UFDs and small dSph galaxies, therefore do not put strong constraints on the time-delay distribution of $r$-process events, though some information might be gained from modeling them in a statistical sense. Currently, the available evidence from UFDs and from the small dSph galaxies is thus fully consistent with our scenario of two separate $r$-process sources.

\section{Conclusions}

With the goal to investigate the timescales of the $r$-process, we looked at the chemical abundances and the star formation histories (SFHs) of the Milky Way solar neighborhood and the dwarf satellite galaxies. The time delay of SNIa (Fe) and AGB stars (Ba) relative to ccSN (Mg) results in the values of [Fe/Mg] and [Ba/Mg] to increase with [Fe/H] in all galaxies with SFH $>1$~Gyr. On the contrary, the Milky Way and the Sculptor dSph galaxy have a relatively flat trend of [Eu/Mg] with [Fe/H], which suggests an $r$-process (main production site of Eu) with short timescales. However, the larger dSph galaxies, Sagittarius and Fornax have an unambiguous increase in [Eu/Mg] with [Fe/H], suggestive of a delayed $r$-process. The available observational data in the Local Group can therefore only be explained by (at least) two $r$-process sources: a \textit{quick} site and a \textit{delayed} site. 

The quick source has to have timescales shorter than those of AGB stars and SNIa (i.e. $t_\textsl{quick}\lesssim100$~Myr) to explain the flat trends in the Milky Way and Sculptor. The delayed source has to have characteristic timescales (i.e. a timescale where the majority of events occur) that are longer than the SFH of Sculptor and longer than the time since the peak in star formation rate (SFR) in the solar neighborhood. Therefore we can constrain $t_\textsl{delayed}\gtrsim4$~Gyr. 

The delayed $r$-process source is most likely neutron star mergers (NSMs) since they are confirmed $r$-process sites (e.g.~\citealt{Watson19}). The estimated time delay of the event GW170817, which is currently the only directly observed NSM with a confirmed host galaxy \citep{Pan17,Blanchard17}, is in good agreement with our timescale of $t_\textsl{delayed}\gtrsim4$~Gyr. The quick source is less certain, but is likely associated with massive stars and could therefore be collapsars \citep{Siegel19b,Siegel19a} or magnetohydrodynamic (MHD) driven SN with jets and strong magnetic fields \citep{Winteler12}. Another fast-track channel of compact object mergers is also a possibility.

The flatness of [Eu/Mg] with [Fe/H] in the Milky Way and the Sculptor dSph suggests that the quick source is the dominant $r$-process site in these galaxies. In general, the quick source is expected to be dominant in systems that have either a uniquely old stellar population (Sculptor) or an increasing SFH over cosmic time (the Milky Way), in other words, systems where the chemical enrichment of each population is dominated by stars $\lesssim4$~Gyr younger. The chemical contribution of the delayed source is only unambiguous in the larger dSph galaxies, Sagittarius and Fornax. This suggests that galaxies that are dominated by an old stellar population \textit{and} have extended SFHs ($>4$~Gyr) are most likely to show the chemical signatures of neutron star mergers.

Our proposed scenario of two $r$-process sources, a quick and a delayed source, is able to self-consistently explain the chemical abundance pattern observed in the Milky Way and all its dwarf satellite galaxies. Furthermore, it is supported by measurements of [Eu/Mg] in solar twins with high-precision astroseismic age measurements. 

Our simple and schematic approach can (and should be) extended to include detailed chemical evolution models of the Milky Way, its different Galactic components, and its dwarf galaxy satellites. By combining self-consistent chemical evolution models with the large amount of available and upcoming data (e.g.~\citealt{Dalton16,Jonsson18,Buder18,deJong19}), we will be able to further test our proposed scenario and put stronger constraints on the complicated timescales of the $r$-process.

\begin{acknowledgements}
The authors would like to thank Mayte Alfaro Cuello, Moritz Reichert  and Thomas de Boer for readily sharing their results with us. This project has received funding from the European Research Council (ERC) under the European Union's Horizon 2020 research and innovation programme (grant agreement No. 804240).
\end{acknowledgements}

\bibliography{heimildir}

\end{document}